# Mixed models for longitudinal left-censored repeated measures

Rodolphe Thiébaut[a*], Hélène Jacqmin-Gadda[a]

[a] INSERM E0338 Biostatistics, ISPED, Université Victor Segalen Bordeaux II, 146, rue Léo Saignat 33076, Bordeaux Cedex, France

## Abstract

Longitudinal studies could be complicated by left-censored repeated measures. For example, in Human Immunodeficiency Virus infection, there is a detection limit of the assay used to quantify the plasma viral load. Simple imputation of the limit of the detection or of half of this limit for left-censored measures biases estimations and their standard errors. In this paper, we review two likelihood-based methods proposed to handle left-censoring of the outcome in linear mixed model. We show how to fit these models using SAS® Proc NLMIXED and we compare this tool with other programs. Indications and limitations of the programs are discussed and an example in the field of HIV infection is shown.

* Corresponding author. Tel: +33 5 57 57 45 21 Fax: +33 5 56 24 00 81

Email: rodolphe.thiebaut@isped.u-bordeaux2.fr

# 1. Introduction

Designs with repeated or grouped measures are common in epidemiological studies. The linear mixed model for correlated gaussian response is increasingly used, especially since availability of methods to fit such models in standard statistical packages like S-PLUS, BUGS or SAS [1-3].

However, longitudinal or grouped data could be complicated by left-censoring of some measures because of a detection limit of the assay used to quantify the marker. For example, this can occur with the concentration of some pollutants in environmental data [4], with antibody titre [5] or with Human Immunodeficiency Virus viral load in blood compartment (HIV RNA) [6, 7]. This latter example will be used throughout this paper because most of the methods to deal with left-censoring through longitudinal models have been proposed in this topic. HIV RNA ranged from 0 (in theory) to 6 $\log_{10}$ copies/ml. The detection limit depends on the assay generation ranging from 4 $\log_{10}$ copies/ml for the first assays available (in 1996) to 0.7 copies/ml today. Despite the improvement of the sensibility of assay, left-censoring is still an issue because antiretroviral treatments available since 1996 are very effective and lead to a steep decrease of HIV RNA after their initiation.

Several methods have been proposed to handle left-censoring of HIV RNA [6-11] rather than imputing the value of the detection limit or half of this limit. For mixed models, one can distinguish multiple imputation [8] and likelihood-based methods [6, 7, 10]. Among likelihood methods, Hughes [6] and Jacqmin-Gadda et al. [7] approaches differ only in the optimisation algorithm used to obtain maximum likelihood estimates. Lyles et al. have published an approach to deal with left-censoring of HIV RNA as well as informative dropout based on an hierarchical formulation of the likelihood [10]. Basically, all these methods lead to less biased estimates and increased standard errors of estimates compared to simple imputation methods. However, the diffusion of these methods could be limited by the use of Fortran [6, 7] or SAS® IML programs [10].

In this paper, we show how to use the new SAS® procedure NLMIXED [12] to fit mixed models taking into account left-censored repeated measures. We compare this procedure based on the conditional formulation of the likelihood given random effects with the algorithm proposed by Jacqmin-Gadda et al. [7] and based on a conditional formulation of the likelihood given observed measures.

## 2. Methods

### 2.1 Model

We considered a linear mixed model [13] applied for modelling $\log_{10}$ HIV RNA. Let $Y_{ij}$, the $j^{th}$ measure at the time $t_{ij}$ $(j = 1,\ldots,n_i)$ for the subject $i$ $(i = 1,\ldots,N)$. For each subject $i$, we distinguished the $n_i^o$-vector of observed response $Y_i^o$ and the $n_i^c$-vector of censored response $Y_i^c$. A general formulation of the linear mixed model with $p$ explicative variables is:

$$Y_i = X_i\beta + Z_i\gamma_i + e_i \text{ with } \begin{cases} e_i \sim N(0, \sigma_e^2 I_{n_i}) \\ \gamma_i \sim N(0, G) \end{cases}$$

where $X_i$ is a $n_i \times p$ design matrix, $\beta$ is a $p$-vector of fixed effects, $Z_i$ is a $n_i \times q$ design matrix which is usually a subset of $X_i$, $\gamma_i$ is a $q$-vector of individual random effects with $q \leq p$. Random effects and measurement errors $(e_i)$ are assumed to be independent.

### 2.2 Likelihood

Basically, two approaches are distinguished according to the development of the likelihood.

In Hughes [6] and Jacqmin-Gadda et al. [7] papers, the likelihood is formulated given $Y_i^o$:

$$L(\theta) = \prod_{i=1}^{N} f_{Y_i^o|\theta}(Y_i^o|\theta) \Pr(Y_i^c < c_i | Y_i^o, \theta)$$

$$= \prod_{i=1}^{N} f_{Y_i^o|\theta}(Y_i^o|\theta) \int_{H_1}\int_{H_2}\ldots\int_{H_{n_i^c}} f_{Y_i^c|Y_i^o,\theta}(u|Y_i^o,\theta) du_1 du_2 \ldots du_{n_i^c} \quad (1)$$

with $\theta$ the vector of model parameters, $u$ a $n_i^c$-vector $[u_1, u_2 \ldots u_{n_i^c}]^T$ and $H_d = ]-\infty, c_{id}]$ the range of integration with $d = 1, 2, \ldots, n_i^c$ and $c_i$ the $n_i^c$-vector of censoring threshold for the subject i. $f_{Y_i^o|\theta}(.|.)$ is the multivariate normal density of observed measures and $f_{Y_i^c|Y_i^o,\theta}(.|.)$ is the multivariate normal conditional density of censored measures given observed measures. Thus, computation of this likelihood needs calculation of a multiple integral as large as the number of censored measured bysubject.

To use the NLMIXED procedure or the Lyles et al. program [10], the likelihood is formulated given the random effects:

$$L(\theta) = \prod_{i=1}^{N} \left[ \int_{R^q} \left\{ \prod_{j=1}^{n_{io}} f_{Y_{ij}^o|\gamma_i}(Y_{ij}^o|\gamma_i = u) \right\} \left\{ \prod_{j=n_{io}+1}^{n_{ic}} \phi_{Y_{ij}^c|\gamma_i}(Y_{ij}^c|\gamma_i = u) \right\} f_{\gamma_i}(u) \, du_1 \, du_2 \ldots du_q \right] \quad (2)$$

$f_{Y_{ij}^o|\gamma_i}(.|.)$ is the univariate normal conditional density of the observed measure $j$ in subject $i$ given random effects and $\phi_{Y_{ij}^c|\gamma_i}(.|.)$ is the univariate normal cumulative distribution function of the censored measure $j$ in subject $i$ given random effects. With this formulation, computation of this likelihood needs calculation of a multiple integral as large as the number of random effects included in the model.

## 2.3 Computation

Whatever the method used, estimation of model parameters is based on maximum likelihood and need computation of multiple integral. However, the approaches differ in term of dimension of the integrals, numerical methods used to compute those integrals and optimisation algorithm used to maximise the likelihood.

The Fortran program proposed by Jacqmin-Gadda et al. [7] to maximise the likelihood (1) is called CENSAD and is available at http://www.isped.u-bordeaux2.fr. It is based on a Marquardt algorithm [14] that is a Newton-Raphson like algorithm where the diagonal of the Hessian matrix is inflated when adapted. To impose a positive constraint of covariance parameters, a new parameterisation of the model was used in term of squared root of $\sigma_e^2$ and a Cholesky decomposition of the random effects' covariance matrix. Multiple integrals of multivariate normal density as large as the number of censored measures $n_i^c$ were numerically calculated using a subregion adaptative multiple integration method [15]. This integration algorithm was found to be optimal when the size of the integral is less than 10 [16]. The algorithm developed by Hughes [6] aims also to maximise the likelihood (1) but it is based on an EM algorithm and a Gibbs sampler to compute the integral in the E step. Those two algorithms have been previously compared in a simulation study and it has been shown that the MCEM algorithm presents more convergence problems and leads to more biased estimates for some covariance structures [7].

The NLMIXED procedure is available since SAS® version 7 and has been written to fit non linear mixed model [12, 17]. This procedure allows specifying a general form for the conditional distribution of the response variable given the random effects. Thus, a general log likelihood function (given the random effects) can be specified using the option *general* in the statement *model*. Then, the procedure directly maximises the approximate integrated likelihood. Several optimisation algorithms can be used with NLMIXED through the option *technique* such as Conjugate Gradient, Double Dogleg, Nelder-Mead Simplex, classic, Ridge or Quasi Newton-Raphson. By default, NLMIXED performs a Quasi-Newton optimisation. This algorithm works with an approximation of the Hessian matrix and thus does not need calculating second order derivatives, which is computationally intensive. This great choice in algorithms allows using the optimal one according to the kind of data [18]. By default, the computation of the integral over random effects is performed by an adaptive Gaussian quadrature method [19] but other methods such as importance sampling (a Monte Carlo method) may be chosen using the option *methods*. The IML program proposed by Lyles et al. [10] and NLMIXED are very close in their approach. However, the IML program was written for a model including only a random intercept and a random slope and then need to be rewritten for another model. The program uses a simple quadrature method to integrate over random effects that is clearly weaker than the adaptative quadrature implemented in NLMIXED [20]. Moreover, computational time is definitely longer than other methods (about half an hour for the presented application). The advantage of IML is the potential of programming very different models and, for example handling informative dropout as proposed by Lyles et al. [10].

Differences between approaches are summarised in table 1.

# 3. Codes for NLMIXED

## 3.1 Univariate mixed model

The model considered in the following included a fixed $(\alpha)$ and random $(a_i)$ intercept and a fixed $(\beta)$ and random $(b_i)$ slope:

$$Y_{ij} = \alpha + a_i + (\beta + b_i)t_{ij} + e_{ij} \qquad (3)$$

with $G = \begin{bmatrix} \sigma_1^2 & \sigma_{12} \\ \sigma_{12} & \sigma_2^2 \end{bmatrix}$, $e_{ij} \sim N(0, \sigma_e^2)$ and $\begin{bmatrix} a_i \\ b_i \end{bmatrix} \perp e_i$

The code to fit such model accounting for left-censoring of $Y_{ij}$ is:

```
proc nlmixed data=TEST QTOL=1E-6;
parms sigsq1=0.4 sig12=-0.03 sigsq2=0.4 sigsqe=0.2 alpha=3.08
beta=0.43;
bounds sigsq1 sigsq2 sigsqe >= 0;
pi=2*arsin(1);
mu=alpha+beta*TIME+a_i+b_i*TIME;
if OBS=1 then ll=(1/(sqrt(2*pi*sigsqe)))*exp(-(RESPONSE-
mu)**2/(2*sigsqe));
if OBS=0 then ll=probnorm((RESPONSE-mu)/sqrt(sigsqe));
L=log(ll);
model RESPONSE ~ general(L);
random a_i b_i ~ normal([0,0],[sigsq1,sig12,sigsq2])
subject=ID;
```

Model parameters must be declared in the first statement *parms*. Starting values are not mandatory but could be crucial in case of convergence difficulties. The statement *bounds* allow specifying some constraints on parameters. In the following example, variance parameters ($\sigma_1^2$, $\sigma_2^2$ and $\sigma_e^2$) are constrained to be positive. The dependent variable named "RESPONSE" is declared to follow a general log-likelihood that returns the value L in the statement *model*. The programming lines between the statements *bounds* and *model* define this log-likelihood given the random effects. The conditional distribution of $Y_{ij}$ given the random effects is defined by expectancy $\mu = E(Y_{ij}|a_i,b_i) = \alpha + \beta t_{ij} + a_i + b_i t_{ij}$ and the variance $Var(Y_{ij}|a_i,b_i) = \sigma_e^2$. Then, according to the status of $Y_{ij}$, i.e. observed (OBS=1) or left-censored (OBS=0), the contribution to the likelihood is the density or the cumulative distribution function calculated by the function *probnorm* of a univariate normal variable. Because the procedure minimises the log-likelihood, the value of the likelihood contribution for the observation $Y_{ij}$ is log-transformed. The distribution for the random effects is specified in the statement *random*. Today, only normal distribution is available with NLMIXED. When defining the lower triangle of the random effects covariance matrix, some of covariance elements can be set to 0. That could be useful to test a covariance with a likelihood ratio statistics. In the statement *random*, the option *subject=id* defined how the dataset is clustered. Nested random effects are not possible at present.

In case of difficulties for convergence, several options are available apart from changing starting points (statement *parms*) and algorithm (options *technique*, *linesearch*, *update*). For example, the option *optcheck* avoid terminating at a stationary point by checking if the likelihood around the convergence point is not better than the likelihood at convergence. The convergence criteria (based on likelihood function, parameters or gradient) can be modified. Sometimes, difficulties come from numerical calculation of integrals or derivatives. Precision or methods for these calculations are modifiable. For example, derivatives can be approximated using forward (*fd=forward*) or central differences (*fd=central*) and tolerance used to adaptively select the number of quadrature points can be decreased (*qtol=10⁻⁴*, by default).

The positive constraint on variance parameters ($\sigma_1^2$, $\sigma_2^2$ and $\sigma_e^2$) is not enough to insure that the covariance G is positive definite. In the example, this constraint can be respected when formulating the constraint on the correlation coefficient. This leads to the same results in the example presented in results section. The code is modified as follow:

```
proc nlmixed data=TEST QTOL=1E-6;
parms sigsq1=0.44 ro=0.09 sigsq2=0.07 sigsqe=0.18 alpha=3.08
beta=0.43;
bounds -1< ro < 1, sigsq1 sigsq2 sigsqe >= 0;
pi=2*arsin(1);
mu=alpha+beta*TIME+a_i+b_i*TIME;
if OBS=1 then ll=(1/(sqrt(2*pi*sigsqe)))*exp(-(RESPONSE-
mu)**2/(2*sigsqe));
if OBS=0 then ll=probnorm((RESPONSE-mu)/sqrt(sigsqe));
L=log(ll);
sig12=(sigsq1*sigsq2)**0.5*ro;
model RESPONSE ~ general(L);
random a_i b_i ~ normal([0,0],[sigsq1,sig12,sigsq2])
subject=ID;
```

## *3.2 Bivariate mixed model*

Extension to a bivariate random effects model as proposed previously with the procedure MIXED [21] is also possible with NLMIXED. Thus, parameters of a bivariate model accounting for left-censoring of one or both markers can be estimated. The key idea is to distinguish the two markers using an indicator (noted *VAR* in the code). Presentation of dataset must be performed as previously reported [21] with, in addition, an indicator of the censoring status of the measure. The code for a bivariate model including four random effects, i.e. one intercept and one slope for each marker, is presented below. In the programming statement, conditional expectancies given random effects are defined according to the marker:

$$\begin{cases} E(Y_{ij}^1 | a_i^1, b_i^1) = \alpha^1 + \beta^1 t_{ij}^1 + a_i^1 + b_i^1 t_{ij}^1 \\ E(Y_{ij}^2 | a_i^2, b_i^2) = \alpha^2 + \beta^2 t_{ij}^2 + a_i^2 + b_i^2 t_{ij}^2 \end{cases}$$

When the covariance matrix of random effects, defined in the statement *random*, is unstructured, ten variance parameters must be estimated:

$$\begin{bmatrix} a_i^1 \\ b_i^1 \\ a_i^2 \\ b_i^2 \end{bmatrix} \sim N\left[ \begin{pmatrix} 0 \\ 0 \\ 0 \\ 0 \end{pmatrix}, \begin{pmatrix} s1 & & & \\ s2 & s3 & & \\ s4 & s5 & s6 & \\ s7 & s8 & s9 & s10 \end{pmatrix} \right]$$

So, model parameters are estimated with this code:

```
proc nlmixed data=TEST;
parms alpha1=0.1 beta1=0.1 alpha2=0.1 beta2=0.1 s1=0.1 s2=0.1
s3=0.1 s4=0.1 s5=0.1 s6=0.1 s7=0.1 s8=0.1 s9=0.1 s10=0.1
sce1=0.1 sce2=0.1;
bounds s1 s3 s6 s10 sce1 sce2>=0;
pi=2*arsin(1);

if VAR=1 then do;
    mu=alpha1+beta1*T+a_i1+b_i1*T;
    sce=sce1;
end;
else if VAR=2 then do;
    mu=alpha2+beta2*T+a_i2+b_i2*T;
    sce=sce2;
end;

if OBS=1 then L=(1/(sqrt(2*pi*sce)))*exp(-(Y-mu)**2/(2*sce));
if OBS=0 then L=probnorm((Y-mu)/sqrt(sce));

ll=log(L);
model Y ~ general(ll);
random a_i1 a_i2 b_i1 b_i2~
normal([0,0,0,0],[s1,s2,s3,s4,s5,s6,s7,s8,s9,s10]) subject=id;
```

## 4. Example

Estimations of parameters for the model (3) were compared according to the two programs previously described. Estimations from a crude approach using the procedure MIXED where left-censored measures were replaced by the value of the threshold were also presented. We used the data set simulated by Lyles et al. [10]. It is available at http://www.blackwellpublishers.co.uk/rss/Volumes/Cv49p4.htm. The true parameters were N=50, $n_i = 5$, $\alpha = 3$, $\beta = 0.5$, $\sigma_1^2 = 0.5$, $\sigma_2^2 = 0.1$, $\sigma_{12} = -0.1$ according to the notations in section 3.1. In the simulated data set, there were 38/250 (15.2 %) left-censored measures of HIV RNA. Results of parameter estimations are reported in table 2. Crude approach without handling left-censoring of HIV RNA leads to biased estimates: especially the fixed slope is underestimated by 14% and the covariance between intercept and random slope is underestimated by 74% compared to less than 10% with other approaches. Obviously, the standard errors of estimates are underestimated with the crude approach because of the simple imputation of the limit of detection for censored measures. The comparison of crude approaches and methods taking into account left-censoring were performed more formally elsewhere [7]. Results according to the method used to take into account left-censoring were similar. Computation times were several seconds for CENSAD and NLMIXED. Analyses were performed in a Windows environment with Pentium III processor. NLMIXED methods were performed using SAS® 8.2. CENSAD has been compiled using Fortran Powerstation 4.0® with default options. The two methods started with estimations obtained from the crude approach.

## 5. Conclusion

We presented two approaches to fit linear mixed models accounting for left-censoring of the response and we showed with an example that they gave the same results. Thus, in a practical point of view, to fit mixed models for left-censored repeated measures, one can choose between NLMIXED and CENSAD. The main elements to choose between approaches are the structure of the data and the model used. In fact, CENSAD will be limited when numerous measures are censored while too many random effects will limit NLMIXED because of the numerical integration. Another point is the potential extension of the estimation to more general model. CENSAD allows including a Gaussian process in the error term like a first order auto-regressive process or a Brownian motion. The extension to a bivariate model is direct with NLMIXED (see section 3.2) and possible using the other method [22]). Using NLMIXED, the main limitation is then the number of random effects. In our experience, the procedure was reliable until four random effects leading, for example, to a bivariate model with two intercepts and two random slopes.

Using the approaches presented in this paper, one must keep in mind the limits of the methods. In particular, the models are fully parametric and assume the normality of outcome distribution. Moreover, in case of missing data, estimations are reliable only if the missngness process is not informative [10].

In conclusion, linear mixed models may be estimated accounting for repeated left-censored measures using available tools. In the context of HIV infections, this could be very useful because of the occurrence of longitudinal analyses of HIV RNA evolution and the bias induced by naïve approaches [6, 7, 10].

Table 1. Summary of different presentations to fit a linear mixed model for left-censored repeated gaussian data.

| Approach | Tool | Integration | Optimisation algorithm |
|---|---|---|---|
| Jacqmin-Gadda et al. (CENSAD) [7] | Fortran 77 | - Over censored measures<br>- Subregion adaptative SADMVN [15] | Marquardt |
| Procedure NLMIXED | Procedure SAS® /STAT module | - Over random effects<br>- Adaptative Gaussian quadrature or others* | Quasi-Newton or others* |

* See text: 2.3 Computation

Table 2. Parameters estimation and standard deviation (sd) of a mixed model with one intercept and one slope according to the method used. Simulated data from Lyles et al. [10]: N=50, $n_i = 5$, $\alpha = 3$, $\beta = 0.5$, $\sigma_1^2 = 0.5$, $\sigma_2^2 = 0.1$, $\sigma_{12} = -0.1$, $\sigma_e^2 = 0.2$.

| Method | $\hat{\alpha}$ (sd) | $\hat{\beta}$ (sd) | $\hat{\sigma}_1^2$ (sd) | $\hat{\sigma}_{12}$ (sd) | $\hat{\sigma}_2^2$ (sd) | $\hat{\sigma}_e$ (sd) |
|---|---|---|---|---|---|---|
| Censoring not handled (SAS® Proc MIXED) | 3.08 (0.10) | 0.43 (0.05) | 0.44 (0.11) | -0.026 (0.042) | 0.066 (0.029) | 0.18 (0.02) |
| SAS® Proc NL MIXED | 2.94 (0.13) | 0.51 (0.062) | 0.66 (0.17) | -0.11 (0.066) | 0.089 (0.040) | 0.23 (0.029) |
| CENSAD | 2.94 (0.13) | 0.50 (0.062) | 0.66 (0.17) | -0.11 (0.066) | 0.089 (0.040) | 0.23 (0.029) |